\documentclass[11pt]{amsart}
\usepackage{geometry}                
\usepackage{graphicx}
 \usepackage{epsfig}
\usepackage{amssymb}
\usepackage{epstopdf}
\title{Flavor Dynamics}

\author{Michael Murray for the BRAHMS Collaboration} 

\begin{document}
\maketitle
The purpose of BRAHMS is to survey the dynamics of relativistic heavy ion (as well as pp and d-A) collisions over a very wide range of rapidity and transverse momentum \cite{br_nim}. The sum of these data may give us a glimpse of the initial state of the system, its transverse and longitudinal evolution and how the nature of the system changes with time \cite{br_white}. Here I will concentrate on the origin and dynamics of the light flavors, i.e. the creation and transport of the up, down and strange quarks. The results presented here are certainly not the end of the story. It is my hope that in a few years new detectors  will reveal the rapidity dependence of the charm and bottom quarks \cite{NewDetectors}. 

\section{Overview of particle production}
\begin{figure}[htbp]
\begin{center}
\epsfig{file=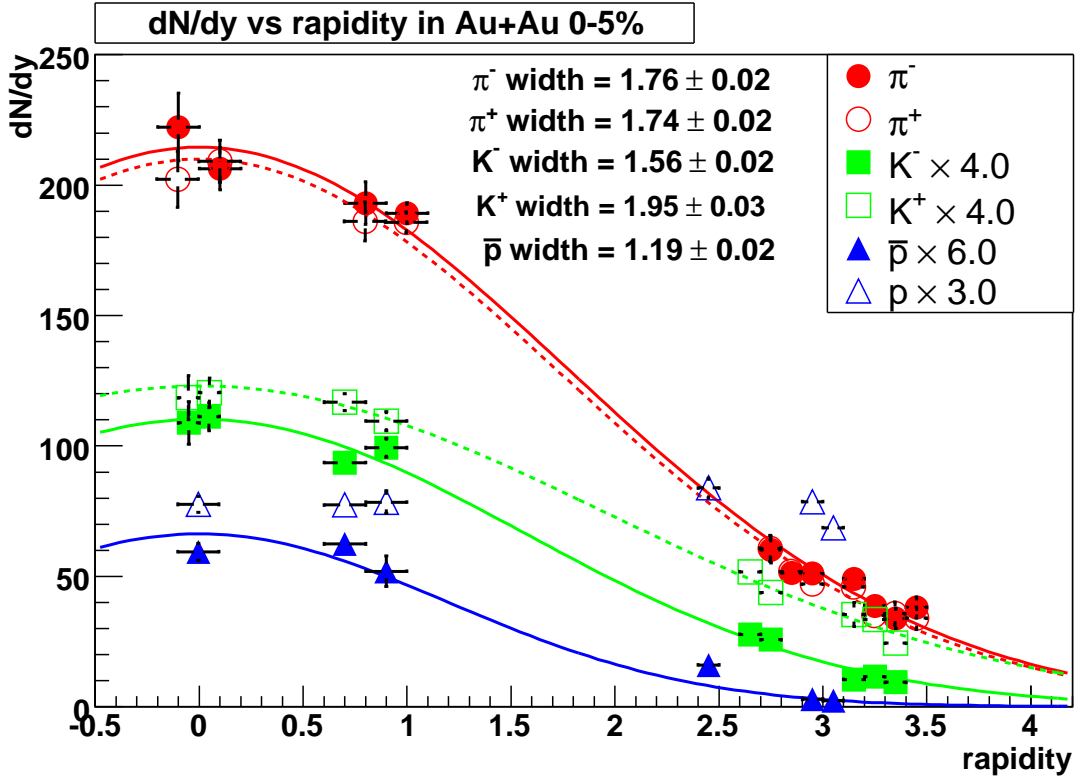,width=\columnwidth}
\caption{Preliminary dN/dy  versus rapidity from central Au+Au collisions at $\sqrt{s_{NN}} = 62.4$GeV }
\label{Yields62GeV}
\end{center}
\end{figure}

Figure \ref{Yields62GeV} shows our preliminary yields of $\pi^\pm, k^\pm$ and antiprotons
versus rapidity for central Au+Au collisions 
at $\sqrt{s_{NN}} = 62.4$GeV. The yields of all produced particles are well described by Gaussian fits up to
 rapidities of $0.8 \times y_{beam}$. The widths of 
 the two pion curves are very close to the prediction by Carruthers for massless particles undergoing Landau
  flow, i.e. $\sigma_\pi = ln{\gamma}$ 
where $\gamma$ is the Lorentz factor of the incoming beams \cite{Landau,Carruthers72,Carruthers73}. This model assumes that all of the entropy of the system is created at the instant the two beams collide and that the system then expands adiabatically until freezeout. One might expect that massive particles would have somewhat narrower widths than this, since for a given amount of energy transferred they will have lower speeds than lighter particles.  The data show  ordering of the rapidity widths that is the same as
 at $\sqrt{s_{NN}} = 200$GeV, namely   $\sigma_{k^+}  > \sigma_{\pi^\pm}  > \sigma_{k^-}  > \sigma_{pbar}$ \cite{BrMeson}.  This is suggestive of a correlation between the production of protons and Kaons. We have investigated this link in terms of a chemical analysis of quark rapidity densities, as discussed below. 

\begin{figure}[htbp]
\begin{center}
\epsfig{file=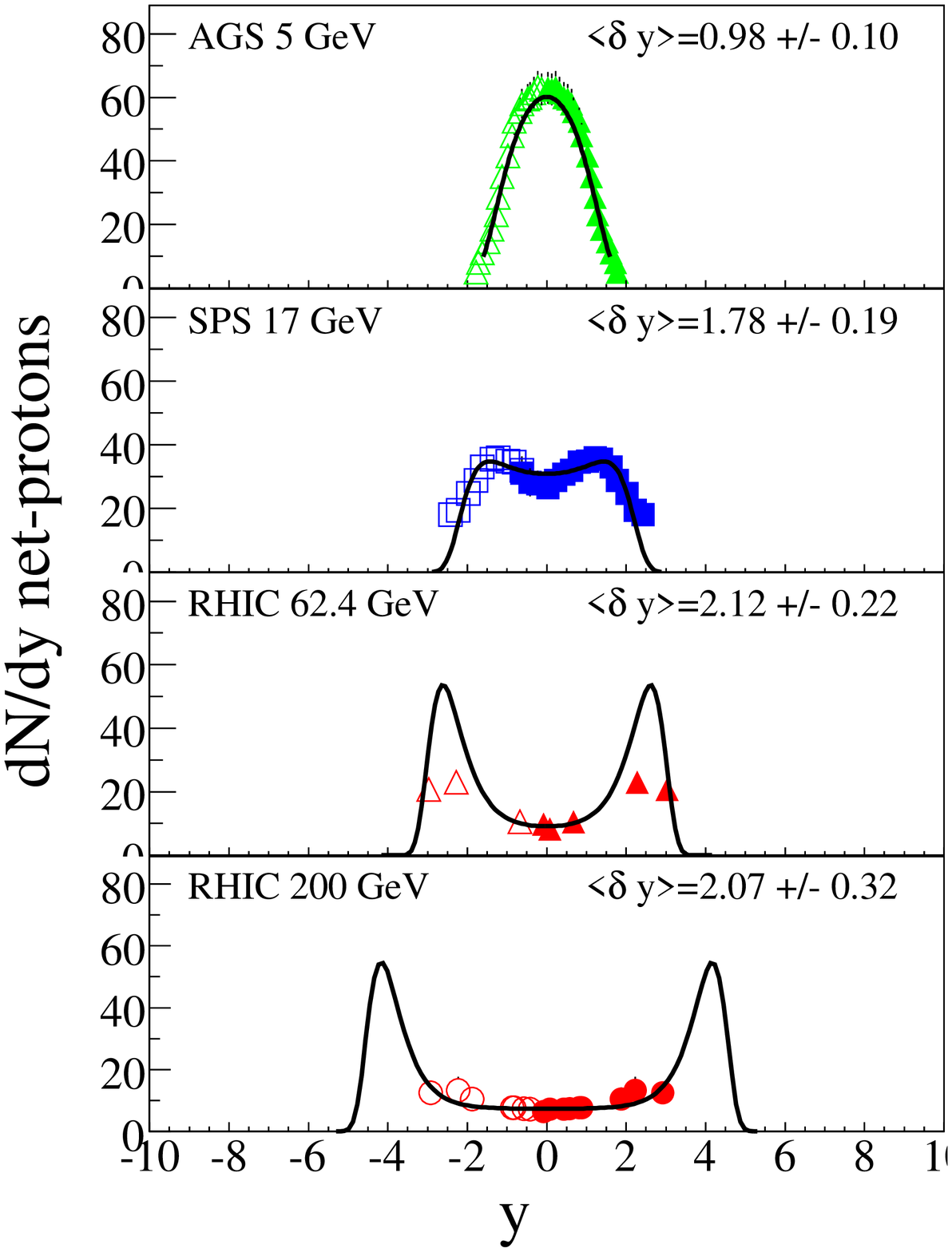,width=\columnwidth}
\caption{\label{NetProt} Net proton distributions versus rapidity and $\sqrt{s_{NN}}$ from E917,E802 and E877 5GeV \cite{E917,E802,E877}, 
NA49 (PbPb) 17GeV \cite{NA49} and  BRAHMS 62 and 200GeV \cite{BrStopping}. The 62GeV data are preliminary with statistical errors only.}
\end{center}
\end{figure}

Since the initial protons are explicity {\it excluded} from the Landau/Carruthers model we should not expect them to be controlled by the same dynamics. Figure \ref{NetProt} shows our distribution of net protons, $dN^p/dy - dN^{\bar p}/dy$ at various energies \cite{BrQM06}. These data suggest that the average rapidity loss $\langle \Delta y \rangle$  of the incoming nucleons rises steadily with the beam rapidity up to $\sqrt{s_{NN}} = 62$ GeV before saturating at  $\langle \Delta y \rangle = 2$ .   

\begin{figure}[htbp]
\begin{center}
\epsfig{file=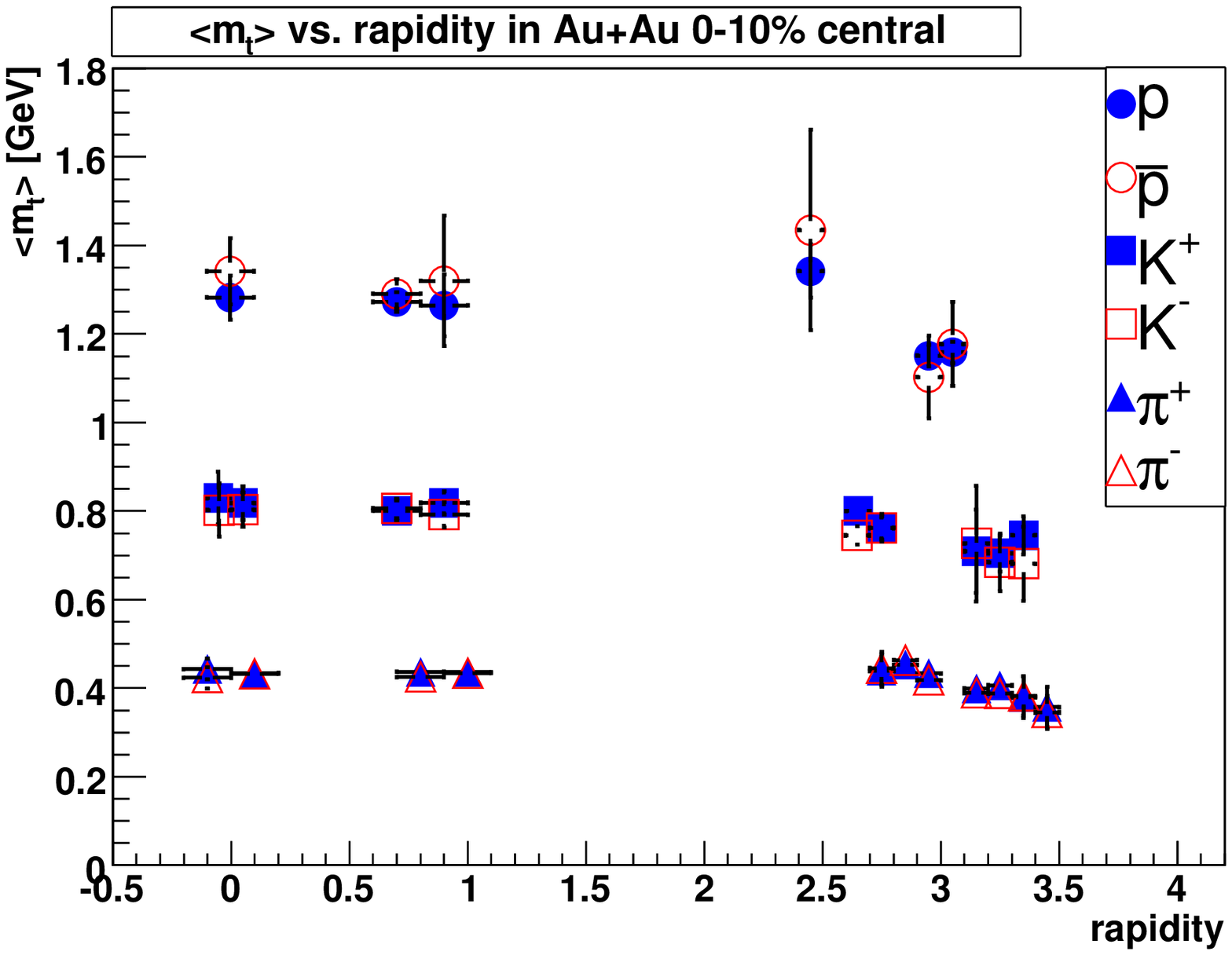,width=\columnwidth}
\caption{Preliminary values of $\langle m_T \rangle$ versus rapidity for  $\pi^\pm, k^\pm, p $  and ${\bar p}$
 for central Au+Au collisions 
at $\sqrt{s_{NN}} = 62.4$GeV. The errors are statistical only.}
\label{Mt62GeV}
\end{center}
\end{figure}

Figure~\ref{Mt62GeV} shows average values of $\langle m_T \rangle$ versus rapidity for  $\pi^\pm, k^\pm$, protons and antiprotons
 for central Au+Au collisions 
at $\sqrt{s_{NN}} = 62.4$GeV. 
The $\langle m_T \rangle$ values are about 10\% lower than at 200GeV for all particles at  all rapidities 
but the relative drop with rapidity is very similar (~14\%). 
Radial flow gives a uniform velocity boost to all particles and so a bigger boost in $\langle m_T \rangle$ to the heavier particles. Our data are consistent with a pattern of radial flow that weakens as the rapidity increases and/or the beam energy drops. 

\begin{figure}[htbp]
\begin{center}
\epsfig{file=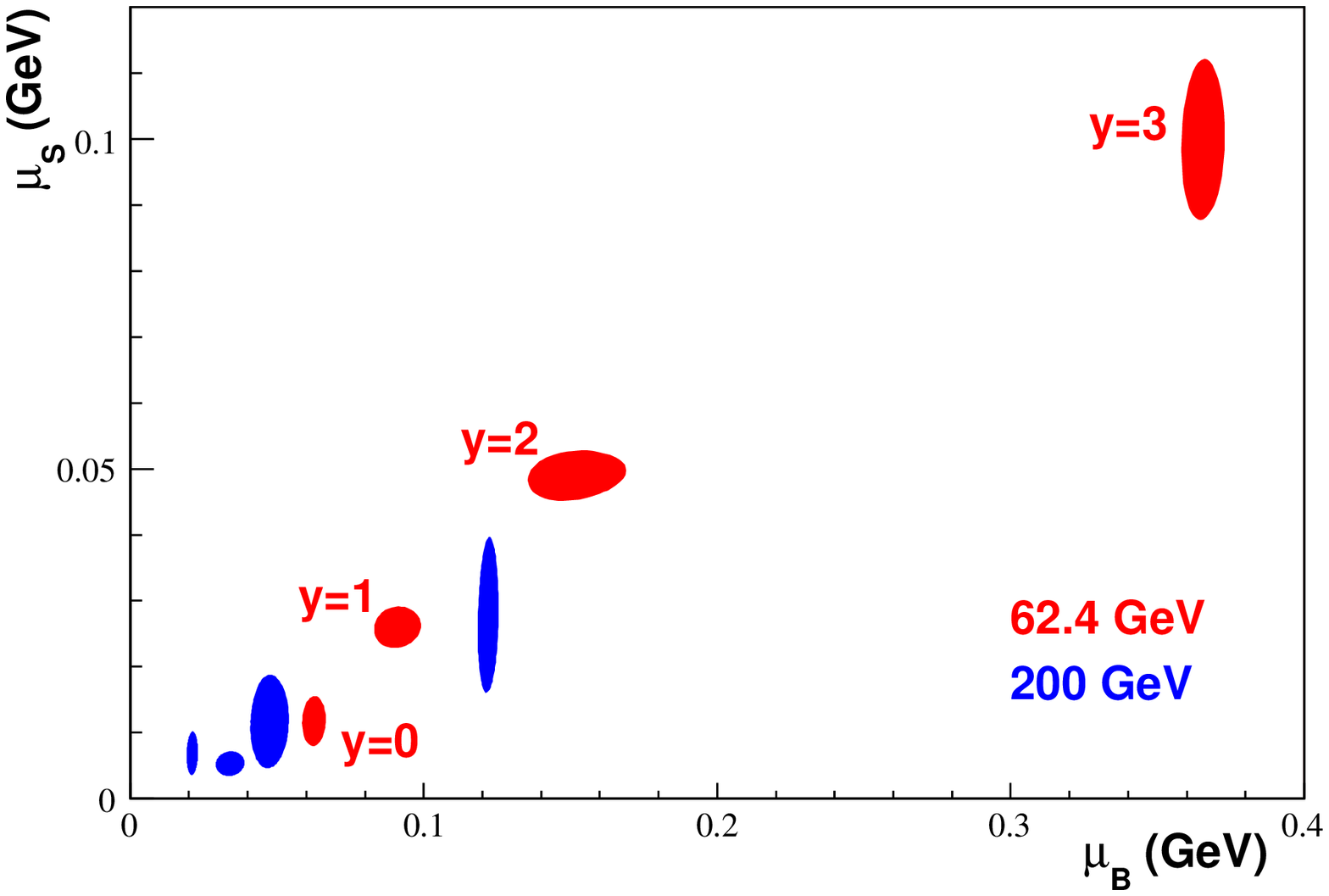,width=\columnwidth}
\caption{Preliminary $\mu_s$  versus  $\mu_q$ at several different rapidities for $\sqrt{s_{NN}} = 62.4$ and 200 GeV. The errors are statistical only.}
\label{TmuBlob}
\end{center}
\end{figure}

The fact that the $k^+$ dN/dy distribution is wider than the pion distribution may be because strangeness is related to the net baryon density. We have investigated this in the context of thermal fits to the data. Using the ``Thermus package" \cite{Thermus}  we have fit five independent particle ratios at each rapidity  to a grand canonical distribution described by a temperature, and two chemical potentials, $\mu_s$ for strange quarks and $\mu_q$ for light quarks.  More details can be found in \cite{Stiles:2006sa}.  
Figure \ref{TmuBlob}  shows $\mu_s$ for versus  $\mu_q$ at several different rapidities for $\sqrt{s_{NN}} = 62.4$ and 200 GeV.  
Note that $\mu_s \approx 1/4 \mu_q$. The chemical freeze-out  temperature drops as the  rapidity increases or $\sqrt{s_{NN}}$  decreases. For example T = $160 \pm 5 $ MeV at  $\sqrt{s_{NN}} = 200$ GeV and y=0 and falls to 
$116 \pm 9 $ MeV at  $\sqrt{s_{NN}} = 62.4$ GeV and y=3.  

\begin{figure}[htbp]
\begin{center}
\epsfig{file=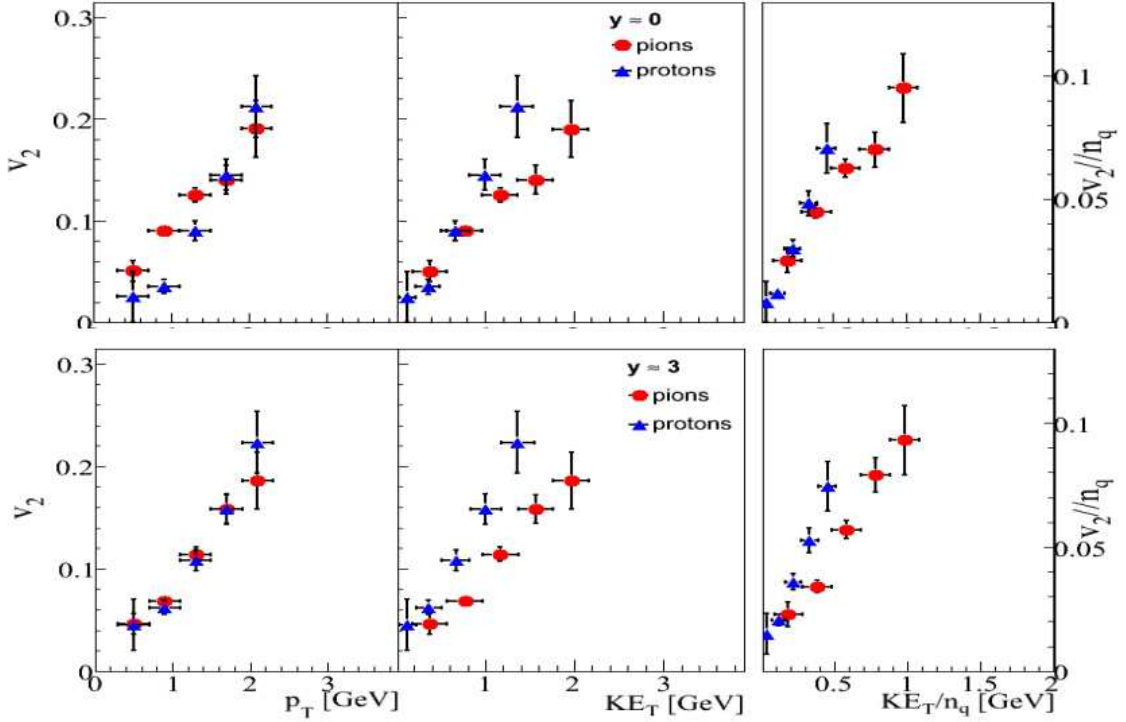,width=\columnwidth}
\caption{Preliminary $v_2$ for protons and $\pi^+$ versus  $p_T$ (left) and $KE_T$ (center) and  $v_2/n_{quark}$ versus $KE_T/n_{quark}$ (right) for  y=0 (top) and y=3 (bottom).  The data are from semi-central  Au+Au collisions at $\sqrt{s_{NN}} = 200$ GeV.  }
\label{V2nquark}
\end{center}
\end{figure}

For non-central heavy ion collisions the geometric eccentricity of the initial state may create pressure gradients if the system is strongly interacting. 
Elliptic flow, i.e. $v_2$  the $2^{nd}$ harmonic of the momentum distribution in azimuthal angle,  is sensitive to these gradients. Since these gradients are 
self quenching $v_2$ tells us about the early state of the system.   At $\sqrt{s_{NN}} =$ 200 GeV we have studied this phenomena as a function of rapidity for pions, Kaons and protons. 
Figure \ref{V2nquark} shows $v_2$ for protons and pions (for clarity the Kaons are not shown) as a function of $p_T$, and $KE_T=\sqrt{p_T^2 + m^2}- m $ as well as $v_2/n_{quark}$ versus $KE_T/n_{quark}$  at central (top) and forward rapidity (bottom). A striking feature of these data is how little  $v_2(p_T)$ changes with rapidity.
This is despite the fact that  when one averages over  $p_T$,  
$\langle v_2 \rangle$
drops by  about a factor of two from y=0 to y=3 \cite{SandersQM06,PhobosV2}. 
The values of $v_2(p_T)$ at all rapidities are close to the maximum values
allowed by ideal hydrodynamics, under the assumption that all of the initial spatial anisotropy is converted into momentum anisotropy. This effect  implies  very early thermalization. 

Objects produced at forward rapidities will tend to come from the collisions of partons with very unequal momenta.   Such collisions have  
a  $\sqrt{s}$ that is lower than the balanced collisions which produce particles near mid-rapidity, and so are less likely to produce particles at high $p_T$. 
We find that the mean $p_T$ of pions, Kaons and protons drops steadily with rapidity  \cite{BrMeson}.
Thus even though $v_2(p_T)$  is rather constant,  $\langle v_2 \rangle$ drops quickly with rapidity. Of course we should not discount the effect of radial and longitudinal flow but this will have to wait for a more complete analysis.  

At y=0 the proton and pion data are closest when plotted as $v_2/n_{quark}$ versus $k_T/n_{quark}$. This effect has been seen by several experiments and suggests that the quarks themselves flow until they coalescence into hadrons.  This scaling does not work quite so well at y=3 where $v_2/n_{quark}$ at a given $k_T/n_{quark}$ is slightly lower for pions than for protons. Since only $  \approx 4\%$  
of our reconstructed pions come from $k^0_{short}$ decays it is unlikely that this effect is caused by a ``dilution" of $\langle v_2 \rangle$ for the pions.  Rather it may reflect a breakdown of coalescence, perhaps caused by recombination, in the forward direction. 

The success of non viscous hydrodynamics in describing both the longitudinal and elliptic flow suggests that thermalization is archived very rapidly in heavy ion collisions.  Black holes are a possible generator of very rapid thermalization \cite{Castorina:2007eb}. Because of the uncertainty principle black holes radiate with a temperature that only depends upon their mass, charge and angular momentum. For uncharged black holes with zero angular momentum ,$T = 1/(8\pi GM)$ where M is the mass, and G is the gravitational constant. The radiation carries away energy which causes the temperature to increase, further increasing the rate at which mass is lost. This effect means that very small black holes would evaporate very quickly, converting there energy to a thermal distribution of particles.  
For charged  black holes the temperature drops, eventually reaching zero when $Q^2 = GM$. One can make  an analogue of strongly coupled  QCD  which can be mapped onto  general relativity using the ADS/CFT correspondence. In this transformation 
\begin{eqnarray*}
M  \rightarrow E, \rm{\ i.e. } \ dE_T/dy \\
Q  \rightarrow B, \rm{baryon \ number} \\
G \rightarrow 1/2 \sigma,  {\rm \  string \ tension} \\
\end{eqnarray*}
and the temperature $T_Q$ of a black hole becomes, 
\begin{equation}
T_Q(B) = T_Q(0) \times \frac{4\sqrt{1-2\sigma B^2/E^2}}{(1+\sqrt{1-2\sigma B^2/E^2})^2}
\label{eq:TempBE}
\end{equation}

which falls very slowly  as the ratio $B/E$ increases until $2\sigma B^2/E^2$ reaches 0.9 when it suddenly dives to zero.
The temperature $T_Q(B)$ is then identified as the chemical temperature deduced from an analysis of particle ratios. 
We make the following  working definitions; 
\begin{eqnarray}
E =  \sum_{i=\pi,k,p} \langle m_T^i  \rangle dN^i/dy \\
 B = dN^p/dy - dN^{\bar p}/dy
\end{eqnarray}
and take  $\sigma = 200 MeV^2$.  

Figures~\ref{Yields62GeV} and ~\ref{Mt62GeV} show that both $\langle m_t \rangle$ and dN/dy decrease with rapidity. At $\sqrt{s_{NN}} = $200 GeV both quantities are larger 
\cite{BrMeson}.  Therefore $dE_T/dy$ decreases with rapidity and increases with $\sqrt{s_{NN}}$. 
Figure \ref{NetProt} shows that  $B=dN^p/dy - dN^{\bar p}/dy$ has the opposite dependence, rising with rapidity and falling with $\sqrt{s_{NN}}$. Thus our largest lever arm for testing Equation~\ref{eq:TempBE} come from comparing T at y=0 and  $\sqrt{s_{NN}}=200 $GeV with T at y=3 and  $\sqrt{s_{NN}}$ = 62.4GeV. 
We find that $2\sigma B^2/E^2$ changes from $(1.0 \pm 0.3) * 10^{-4}$ to $(3.6 \pm 0.2) * 10^{-2}$ over this range while the freeze-out temperature drops from $160 \pm 5 $ MeV to $116 \pm 9$. 
To accommodate such a temperature change within the model would require rescaling our $B/E$ measurements by a factor of 5. 
Using HIJING we estimate that the net baryon number is a factor of 2.0 times the net proton yield at y=0. At y=3 this factor rises to 2.1. Modifying our definition working of $B/E$ to use these estimates and taking $\pi^0 = (\pi^+ + \pi^-)/2$ increases $B/E$ by a factor of $1.4 \pm 0.1$. 

\section*{Conclusions}
The rapidity distributions of particles produced from central Au+Au collisions are Gaussian at RHIC energies with widths that  increase with $\sqrt{s_{NN}}.$ 
The pion widths are well described by the Landau/Caruthers model of  a relativistic fluid expanding at constant entropy in one dimension until it hadronizes into massless particles. Of course the motion of the particles is not only along the z axis. The increases $\langle m_T \rangle$ with mass suggests there is  a radial component to the expansion. The strength of the radial flow decreases with energy and rapidity. The fact that the scaled elliptic flow $v_2(k_T)/n_{quark}$ is proportional to   $k_T/n_{quark}$ at y=0 suggests that it is the partons that flow before they coalesce into hadrons. There are hints that this proportionality breaks down at y=3 but we need to finish our study of the systematic errors of this analysis.   
The ordering of the rapidity widths, $\sigma_{k^+}  > \sigma_{\pi^\pm}  > \sigma_{k^-}  > \sigma_{pbar}$ does not depend upon energy and has been investigated in terms of a thermal model of flavor production. The chemical freeze-out temperature drops with rapidity and energy while
$\mu_s \approx 1/4 \mu_q$. All of these data are suggestive of very early thermalization. This can be explained by assuming that particles are radiated from black holes. We have made the first moves towards testing this hypothesis. Currently it seems that this theory predicts too slow a change of 
the chemical freeze-out temperature as the ratio of net baryon number to transverse energy increases.


\begin{thebibliography}{12}

\bibitem{br_nim} 
  M.~Adamczyk {\it et al.} [BRAHMS Collaboration],
  Nucl.\ Instr.\ and Meth.\ A 499 (2003) 437.
  
\bibitem{br_white} 
  I.~Arsene {\it et al.} [BRAHMS Collaboration],
  Nucl.\ Phys.\ A 757 (2005) 1.

\bibitem{NewDetectors} For an overview of the heavy flavor capabilities of future experiments see the proceedings of ``Quark Matter 2006", 
F. Antinori {\it et al.} [ALICE Collaboration], J. Phys. G, {\bf 34}, S511; //
R. R. Betts {\it et al.}  [CMS Collaboration], J. Phys. G, {\bf 34}, S519; //
P Steinberg  {\it et al.} [ATLAS Collaboration],  J. Phys. G, {\bf 34} S527; //
 A Rose {\it et al.}  [STAR Collaboration] J. Phys. G, {\bf 34},  S517; //
 A Milov  {\it et al.}  [PHENIX Collaboration] J. Phys. G, {\bf 34} S701 (2007). 

 
 
\bibitem{Landau} L. D. Landau, Izv. Akad. Nauk SSSR, Ser. Fiz. 17, 51 (1953).
 \bibitem{Carruthers72}  P. Carruthers and M. Duong-van, Phys. Lett. B 41, 597 (1972). 
\bibitem{Carruthers73} P. Carruthers and M. Duong-van, Phys. Rev. D 8, 859 (1973).


\bibitem{BrMeson}
I. G. Bearden {\it et al.}  [BRAHMS Collaboration], 
Phys. Rev. Lett. 94 (2005) 162301. 
\bibitem{BrQM06} I. G. Bearden {\it et al.}  [BRAHMS Collaboration], 
J. Phys. G, {\bf 34} S207 (2007)

\bibitem{E917}  B. B. Back {\it et al.,}  [E917 Collaboration], Phys. Rev. Lett. {\bf 86},
1970 (2001).
\bibitem{E802} L. Ahle {\it et al.,}  [E802 Collaboration],  Phys. Rev. C {\bf 60},
064901 (1999).
\bibitem{E877}  J. Barette {\it et al.,} [E877 Collaboration], Phys. Rev. C {\bf 62},
024901 (2000).
\bibitem{NA49} H. Appelshauser {\it et al.,}   [NA49 Collaboration], Phys. Rev.
Lett. {\bf 82}, 2471 (1999).

\bibitem{BrStopping} 
 I. G. Bearden {\it et al.}  [BRAHMS Collaboration], 
Phys. Rev. Lett. {\bf 93}, 102301 (2004)


\bibitem{Thermus} S. Wheaton, J. Cleymans J. Phys. G {\bf 31} S1069, (2005) and  
S. Wheaton and J. Cleymans, hep-ph/0407174

\bibitem{Stiles:2006sa}
  L.~A.~Stiles and M.~Murray,
  arXiv:nucl-ex/0601039.

\bibitem{SandersQM06} S. J. Sanders {\it et al.}  [BRAHMS Collaboration], 
J. Phys. G, {\bf 34} S1083 (2007)
\bibitem{PhobosV2} B. B. Back {\it et al.}  [PHOBOS Collaboration],
Phys. Rev. Lett. {\bf 94}, 122303 (2005)


\bibitem{Castorina:2007eb}
  P.~Castorina, D.~Kharzeev and H.~Satz,
  Eur.\ Phys.\ J.\  C {\bf 52}, 187 (2007)
  [arXiv:0704.1426 [hep-ph]].
\end{thebibliography}
\end{document}